\documentclass[twocolumn,showpacs,preprintnumbers,amsmath,amssymb]{revtex4-2}
\usepackage{graphicx}
\usepackage{dcolumn}
\usepackage{bm}
\usepackage{xcolor}
\usepackage{multirow}
\usepackage{epsfig}
\usepackage{epstopdf}
\usepackage{amsmath}

\begin{document}
	
	
	\title{Complete Next-to-Next-to-Leading-Order QCD Correction to $J/\psi \to 3\gamma$ Decay}
	
	\author{Chao Zeng$^{1,2}$}
	\author{Bin Gong$^{1,2}$}
	\author{Jian-Xiong Wang$^{1,2}$}
	\author{Ruichang Niu$^{1,2}$}
	\author{Xu-Dong Huang$^{3}$}
		\author{Cong Li$^{4}$}
	\affiliation{Institute of High Energy Physics, Chinese Academy of Sciences, Beijing, 100049, P.R. China}
	\affiliation{University of Chinese Academy of Sciences, Chinese Academy of Sciences, Beijing, 100049, P.R. China}
	\affiliation{College of Physics and Electronic Engineering, Chongqing Normal University, Chongqing 401331, P.R. China} 
	\affiliation{School of Physical Science and Technology, Southwest University, Chongqing, 400700, P.R. China}
	
	\date{\today}
	
\begin{abstract}
	We address the long-standing problem of negative decay and production rates in perturbative QCD for exclusive processes by proposing amplitude-level NRQCD factorization as a systematic prescription. Building on this, we present the first complete next-to-next-to-leading-order (NNLO) QCD correction to the decay $J/\psi \to 3\gamma$. The resulting partial width, $\Gamma(J/\psi \to 3\gamma) = 0.96^{+4.32}_{-0.13}$ eV, combines this NNLO contribution with the known up to $\mathcal{O}(\alpha_s v^2)$ relativistic correction and shows markedly improved agreement with the high-precision BESIII measurement. In the same way, $\Gamma(\Upsilon \to 3\gamma) = 0.0086^{+0.0028}_{-0.0006}$ eV is obtained. The dominant theoretical uncertainty originates from the renormalization scale variation, underscoring the challenge of perturbative convergence at this order and the necessity for future higher-order calculations.
\end{abstract}
	
	
	\maketitle
	
	The study of heavy quarkonium annihilation decays has historically served as a crucial proving ground for our understanding of the strong interaction, playing an instrumental role in the early establishment of asymptotic freedom in Quantum Chromodynamics (QCD) \cite{Appelquist:1975, DeRujula:1975}. The systematic and rigorous treatment of these bound-state processes was later revolutionized by the development of the nonrelativistic QCD (NRQCD) effective field theory \cite{Bodwin:1994jh, Caswell:1986}, which provides a formal factorization framework to separate perturbative short-distance physics from nonperturbative long-distance effects. {\color{blue}Among various theoretical approaches, NRQCD currently stands as the most effective theoretical framework for decay or production processes of heavy quarkonium.}

	Within this framework, the exclusive electromagnetic decay $J/\psi \to 3\gamma$ emerges as an exceptionally pristine and theoretically valuable process. As the QCD analogue of the classic QED process orthopositronium $\to 3\gamma$, whose precise theoretical description played a pivotal role in validating the bound-state formalism of NRQED \cite{Caswell:1976nx, Adkins:1996}, the $J/\psi \to 3\gamma$ decay offers a unique and sensitive probe into the interplay between perturbative QCD corrections and the nonperturbative, relativistic dynamics inside the charmonium bound state. Its decay width is thus a benchmark observable for testing the convergence and predictive power of the NRQCD factorization program at an increasingly sophisticated level.
	
	The CLEO-c collaboration first observed this rare decay in 2008, reporting ${\rm Br}(J/\psi\to 3\gamma)= (1.2\pm 0.3\pm 0.2)\times 10^{-5}$~\cite{CLEO:2008qfy}. The BESIII collaboration later refined the measurement to ${\rm Br}(J/\psi\to 3\gamma)=(11.3\pm 1.8\pm 2.0)\times 10^{-6}$~\cite{BESIII:2012lxx}. These increasingly precise results have posed a persistent challenge to theory.

	{\color{blue}On the theory side, }the leading-order (LO) NRQCD prediction exceeds these experimental values by a factor of several. The next-to-leading-order (NLO) QCD correction \cite{Caswell:1976nx, Mackenzie:1981sf} was expected to reduce the gap, but turned out to be large and negative, moving the prediction further from the data. Similarly, the leading $\mathcal{O}(v^2)$ relativistic correction \cite{Keung:1982jb}, accounting for the finite charm quark relative velocity, also contributed a large negative shift. The combination of these two corrections led to severe deterioration, even yielding non-physical negative rates—a long-standing "sign crisis" in the theoretical description of this decay. The perturbative convergence is clearly unsatisfactory. In the large-$N_f$ limit, the series was estimated to be badly divergent due to the large residue of the $u=1/2$ infrared renormalon, especially for $J/\psi \rightarrow 3 \gamma$~\cite{Braaten:1998au,Bodwin:1998mn}.

        {\color{blue}Subsequently, the joint $\mathcal{O}(\alpha_s v^2)$ correction \cite{Feng:2012by} was found to be large and positive, counterbalancing the negative NLO and $\mathcal{O}(v^2)$ contributions. With a natural choice of renormalization scale, the theoretical prediction agreed with both CLEO-c and BESIII data, resolving the sign crisis and validating the NRQCD approach at that order. 
        However in a later analysis \cite{Sang:2020zdv}, it is found  that prediction for the decay width becomes much smaller than the experimental measurements when the data for $\mathrm{Br}(J/\psi\rightarrow \mathrm{LH})$ is used as input instead of NRQCD long-distance matrix elements. The theoretical predictions based on NRQCD for this decay seem unstable, suffering bad convergence and large uncertainty. This leads to immediate questions: How large is the next-order correction in both $\alpha_s$ and $v^2$? Will it make the decay width negative again? Is there a systematic way to handle such issues in perturbative QCD expansions?
        
        	In 2019, a lattice QCD study gave the branching fraction as $(2.13\pm 0.14 \pm 0.29)\times 10^{-5}$~\cite{Meng:2019lkt}, which  is about twice as large as the experimental measurements. Although the light-by-light contribution (where two photons couple through a virtual light-quark loop), which may improve the result, is not included in this analysis, we can conclude that there is still no satisfying theoretical predictions for this decay at present, no matter from perturbative or non-perturbative QCD.}

	In this Letter, we address these questions by computing the next-order QCD correction to $J/\psi \to 3\gamma$—the next-to-next-to-leading-order (NNLO) QCD correction.  {\color{blue}Since the initial particle $J/\psi$ is a $c\bar{c}$ bound state, this is a quasi-$2\to 3$ two-loop calculation}. This requires pushing the frontier of loop calculations with available computational tools. We propose a systematic prescription to avoid negative decay or production rates in perturbative QCD expansions for exclusive processes, and present a state-of-the-art theoretical prediction with significantly improved agreement with the latest BESIII data.
	
	Before describing the calculation, we present the results for $\Gamma(J/\psi \rightarrow 3 \gamma)$:
	\begin{equation}\label{key}
		\Gamma=\Gamma_{LO}\left(1-12.6 \frac{\alpha_s}{\pi} - 8.23\langle v^2 \rangle  + 68.9 \frac{\alpha_s}{\pi} \langle v^2 \rangle  - 28.7 \frac{ \alpha^2_s}{\pi^2} \right),
	\end{equation} 
	where the $\alpha^2_s$ term is obtained for the first time, and input parameters $\mu_R,\mu_\Lambda,m_c$ follow \cite{Feng:2012by}.  With $\alpha_s(m_c)=0.388$ and $v^2=0.225$ as in \cite{Feng:2012by}, the NNLO decay rate in Eq.(1) unfortunately becomes large and negative again. Without this new term, the result in \cite{Feng:2012by} agreed with experiment. The large negative NLO results for $\alpha_s$ and $v^2$ were obtained over 40 years ago and have been cited extensively without a systematic prescription. 
	{\color{blue}Even for $\Gamma(\Upsilon \rightarrow 3 \gamma)$, where the perturbative expansion convergence is much better than it for $J/\psi$, the NNLO result becomes negative.}
	Negative production rates also appeared in $e^+e^-\rightarrow J/\psi + J/\psi$ at QCD NLO~\cite{Gong:2008ce} and NNLO~\cite{Sang:2023liy,Huang:2023pmn}, indicating poor perturbative convergence in these cases.
	
	Factorization for exclusive processes must be implemented at the amplitude level. Therefore, fixed-order QCD corrections should be presented in an amplitude-level factorized form. This ensures that negative decay or production rates never appear in perturbative calculations for exclusive processes at any order.
	
	For $J/\psi\rightarrow 3 \gamma$, NRQCD factorization at the amplitude level gives the decay rate as 
	\begin{equation}
		\Gamma_M=\int d\phi (J/\psi \rightarrow 3 \gamma)\sum_{\mathrm{pol}} |\sum_{l,i}\mathcal{A}_i^{(l)} \langle 0| \hat O_i|J/\psi\rangle |^2, \label{eqn:factornization}
	\end{equation}
	where $\langle 0| \hat O_i|J/\psi\rangle$ is the $i$-th color-singlet long-distance matrix element (LDME), and $\mathcal{A}_i^{(l)}$ is the corresponding short-distance coefficient (SDC) at $l$-th order in $\alpha_s$. Here $M$ = LO, NLO, NNLO, $\dots$ corresponds to summing $l$ up to 0, 1, 2, $\dots$. The expansion is truncated at fixed order at the amplitude level, rather than at the amplitude-square level. This formula can be generalized to any exclusive process as a systematic prescription.
	
	A similar amplitude-level NRQCD factorization was used in the discussion of SDC expansion convergence for $J/\psi\rightarrow 3 \gamma$ in Eq.(126) of Ref.~\cite{Braaten:1998au}.
	
	We refer to Eq.(\ref{eqn:factornization}) as amplitude expansion, and the traditional approach as amplitude-square expansion. {\color{blue}The amplitude-level factorization for exclusive processes means that all the initial and final particles are colorless for QCD corrections, and are electrically chargeless for QED corrections. Therefore no real IR correction is needed to cancel the IR divergence at the amplitude-square level.} Our calculation follows this prescription.
	
	Only the leading LDME $\langle 0| \chi^+ \sigma \psi|J/\psi\rangle$ in the $v^2$ expansion is considered, where $v$ is the charm quark velocity in the $J/\psi$ rest frame. 
	Dimensional regularization with $D=4-2\varepsilon$ is used for both UV and IR divergences.  
	The NNLO renormalization constant for the corresponding operator, from Refs.~\cite{Beneke:1997jm,Kniehl:2006qw}, is 
	\begin{equation}
		Z=\left[1-\alpha_s^2\left(\frac{\mu_{\Lambda}^2e^{\gamma_E}}{\mu_R^2 4\pi}\right)^{-2\epsilon}\left(\frac{C_F^2}{3}+\frac{C_F C_A}{2}\right)\frac{1}{4\epsilon}\right].
	\end{equation}
	The heavy quark field and mass are renormalized in the on-shell (OS) scheme. The coupling $\alpha_s$ is renormalized in the $\overline{\rm MS}$ scheme.
	
	To compute at amplitude level, we project the amplitude onto a complete set of Lorentz structures. The set from Ref.~\cite{Adkins:1996odo} is not sufficiently efficient for the two-loop calculation. We obtain a complete basis of 14 Lorentz structures:
	\begin{eqnarray}\label{eqn:basis}
		\epsilon_{1,2,3}&=&(\varepsilon\cdot\varepsilon_1)(\varepsilon_2\cdot\varepsilon_3),\hspace{0.4em}(\varepsilon\cdot\varepsilon_2)(\varepsilon_1\cdot\varepsilon_3),\hspace{0.4em}(\varepsilon\cdot\varepsilon_3)(\varepsilon_1\cdot\varepsilon_2),  \nonumber\\
		\epsilon_{4,5,6}&=&(\varepsilon\cdot\varepsilon_1)(k_1\cdot\varepsilon_2)(k_2\cdot\varepsilon_3),\hspace{0.4em}(\varepsilon\cdot\varepsilon_2)(k_3\cdot\varepsilon_1)(k_2\cdot\varepsilon_3), \nonumber\\
		&&(\varepsilon\cdot\varepsilon_3)(k_3\cdot\varepsilon_1)(k_1\cdot\varepsilon_2),  \nonumber\\
		\epsilon_{7-12}&=&(k_{1,2}\cdot\varepsilon)(\varepsilon_1\cdot\varepsilon_2)(k_2\cdot\varepsilon_3),\hspace{0.4em}(k_{1,2}\cdot\varepsilon)(\varepsilon_2\cdot\varepsilon_3)(k_3\cdot\varepsilon_1), \nonumber\\
		&&(k_{1,2}\cdot\varepsilon)(\varepsilon_1\cdot\varepsilon_3)(k_1\cdot\varepsilon_2),  \\
		\epsilon_{13,14}&=&(k_{1,2}\cdot\varepsilon)(k_3\cdot\varepsilon_1)(k_1\cdot\varepsilon_2)(k_2\cdot\varepsilon_3) .\nonumber
	\end{eqnarray}
	where $P$, $k_{1,2,3}$ and $\varepsilon$, $\varepsilon_{1,2,3}$ are the momenta and physical polarization vectors of $J/\psi$ and the three photons, with the gauge choice $k_2\cdot\varepsilon_1=k_3\cdot\varepsilon_2=k_1\cdot\varepsilon_3=0$. 
	
	The SDCs are then
	\begin{equation}
		\mathcal{A}^{(l)} =c_i^{(l)} \epsilon_i,  \quad i=1,\cdots,14,
	\end{equation}
	where $l=0,1,2$ denotes LO, NLO, NNLO respectively, and $c_i^{(l)}$ is the projection onto each Lorentz structure.
	
	The inner product of two Lorentz structures is defined as
	\begin{equation}
		\langle\epsilon_j, \epsilon_i\rangle\equiv \sum_{\mathrm{pol}} \epsilon_i \epsilon_j^*\equiv G_{ij}. \label{eqn:inner_product}
	\end{equation}
	The sum runs over physical polarizations of all external particles, and $G_{ij}$ is the metric tensor, satisfying $G_{ij}^*=G_{ji}$. 
	
	All terms in the square of the SDC summation are then
	\begin{align}
		&	\langle \mathcal{A}^{(l_1)}, \mathcal{A}^{(l_2)}\rangle = c_i^{(l_2)} G_{ij}c_j^{(l_1)*}  =d_i^{(l_2)} G^{-1}_{ij}d_j^{(l_1)*}, \nonumber \\
		&	d_j^{(l)} \equiv \langle \epsilon_j, \mathcal{A}^{(l)}\rangle =c_i^{(l)}  \langle\epsilon_j, \epsilon_i \rangle=c_i^{(l)} G_{ij}.
	\end{align}
	
	The determinant of $G_{ij}$ is proportional to $\varepsilon^2$, so the basis in Eq.(\ref{eqn:basis}) becomes dependent when $D=4$, and the first 12 structures $\epsilon_{1-12}$ form a complete basis for $D=4$. Since all SDCs are finite, $\epsilon_{13,14}$ are not needed. This is confirmed by checking $\langle \epsilon_{13(14)},\mathcal{A}^{(l)}\rangle$ against $G_{13(14),i}G^{-1}_{ij}d^{(l)}_j$ ($i,j=1,\dots,12$). 
	\begin{figure}
		\centering
		\includegraphics[scale=0.4]{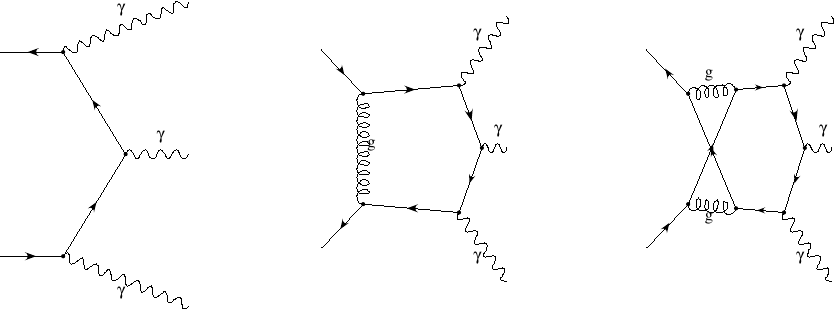}
		\caption[Typical Feynman diagram]{Typical Feynman diagrams for LO, NLO, NNLO.}
	\end{figure}
	
	The calculation of all $d_j^{(l)}$ follows our previous work~\cite{Huang:2023pmn}. First, 6 tree-level, 48 one-loop, and 894 two-loop Feynman diagrams for $J/\psi\rightarrow 3\gamma$ are generated with FeynArts~\cite{Hahn:2000kx}.  
	Lorentz contraction, Dirac and color traces are handled by private code optimized for long fermion chains. 
	The package \texttt{CalcLoop}~\cite{Calcloop} then decomposes all amplitudes into 6 and 96 Feynman integral families for the $A^{(1)}$ and $A^{(2)}$ contributions, respectively. Kira~\cite{Kira:2021} reduces all integrals to master integrals. Finally, AMFlow~\cite{Liu:2022chg} computes all master integrals.
	
	As expected, all coefficients $c^{(l)}_i$ and $d^{(l)}_i$ are finite, confirming NRQCD factorization at QCD NNLO for $J/\psi \to 3\gamma$ at amplitude level. 
	
	For phase space integration, we use the Gauss-Kronrod quadrature rule. The phase space depends on two kinematic invariants $s_1$ and $s_2$. We adopt the G7-K15 rule, discretizing each invariant with 15 points, giving $15 \times 15 = 225$ sampling points across the phase space. {\color{blue}The uncertainty from the phase-space integration is about 0.6\% estimated in G7-K15 algorithm, and verified for the LO and NLO parts against the LO analytic result and NLO results obtained in previous research works~\cite{Mackenzie:1981sf, Campbell:2007ws}.}
	
	At each sampling point, we use the differential equation method, which was first employed in~\cite{Liu:2020kpc}. For each integral family, the target integrals are reduced to master integrals by integration-by-parts (IBP) identities, keeping $s_1$ and $s_2$ symbolic. Differential equations for these master integrals in $s_1$ and $s_2$ are then set up by differentiating the master integrals as per Eq.(8) and subsequently applying IBP reduction to express the results in terms of the master integrals. Both IBP reduction and differential equation construction are done with Kira. High-precision numerical values for these master integrals at a single phase space point are computed with AMFlow as boundary conditions. Using the differential equation solver in AMFlow, the master integrals are evaluated at all remaining sampling points.
	\begin{align}
		&k_l\cdot\frac{\partial I}{\partial k_i}=\frac{\partial I}{\partial (k_i\cdot k_j)}\frac{\partial(k_i\cdot k_j)}{\partial k_i}\cdot k_l, (i,j,l=1,2,3)
	\end{align}
	We obtain positive decay widths at LO, NLO, and NNLO:
	
	\begin{align}
		&\Gamma_{LO}= \frac{16(\pi^2-9)q_c^6\alpha^3|R_{J/\psi}|^2}{3\pi M_{J/\psi}^2}, \nonumber \\  
		&\Gamma_{NLO}= \Gamma_{LO} \Bigl[  1 -12.63 (\alpha_s/\pi) 
		+45.28(\alpha_s/\pi)^2 \Bigr],
		\nonumber \\  
		&\Gamma_{NNLO}= \Gamma_{LO} \Bigl\{1-12.63(\alpha_s/\pi) + (\alpha_s/\pi)^2\bigl[5.369n_l\nonumber\\
		&-1.986 \overline{q}^2-41.77+(4.210n_l-65.26)l_R - 51.18 l_\Lambda\bigr]\nonumber\\
		&+(\alpha_s/\pi)^3\bigl[(467.4-30.16n_l)l_R + 323.2l_\Lambda-42.27n_l\nonumber\\
		&+10.10\overline{q}^2+647.4\bigr]+(\alpha_s/\pi)^4\bigl[654.7l_\Lambda^2+(1670.\nonumber\\
		&-107.7n_l)l_Rl_\Lambda+(1208.-155.9n_l+5.028n_l^2)l_R^2\nonumber\\
		&+ (52.21\overline{q}^2-3.368n_l\overline{q}^2+3346.-434.3n_l+14.10n_l^2)l_R\nonumber\\
		&+(50.82\overline{q}^2+2228.-137.4n_l )l_\Lambda+4.018\overline{q}^4+63.97\overline{q}^2\nonumber\\
		&-3.448n_l\overline{q}^2+2367.-314.0n_l+10.65n_l^2\bigr] \Bigr\}.
		\label{NNLO-result}
	\end{align}
	
	where $l_R=\ln\frac{\mu_R}{m_c}$, $l_\Lambda=\ln\frac{\mu_{\Lambda}}{m_c}$, $q_c=\frac{2}{3}$, $\overline{q}^2=(q_u^2+q_d^2+q_s^2)/q_c^2$ and $n_l=3$ counts the massless quarks. The $2\text{Re}(\mathcal{A}^{(2)}\mathcal{A}^{(0)*}) + |\mathcal{A}^{(1)}|^2$ part is also computed by direct amplitude square, without using the Lorentz basis, as a consistency check.
	
	For numerical results, {\color{blue}the following central values are used}: charm mass $m_c = 1.5\,\text{GeV}$, $J/\psi$ mass $M_{J/\psi}=2m_c$, wave function at origin $|R(0)|^2 = 0.934\,\text{GeV}^3$ from the color-singlet matrix element $|\langle 0| \chi^+ \sigma \psi|J/\psi\rangle|^2$~\cite{Bodwin:2008}, renormalization scale $\mu_R = 2m_c$, NRQCD factorization scale $\mu_{\Lambda}=m_c$, $\langle v^2 \rangle=0.225$~\cite{Bodwin:2008}, and $\alpha_s^{(3)}(\mu_R)=0.25406$ from three-loop running via RunDec3{}~\cite{RunDec:2017}. 
	
\begin{figure}[h]
		\centering
		\includegraphics[width=8cm, height=4cm]{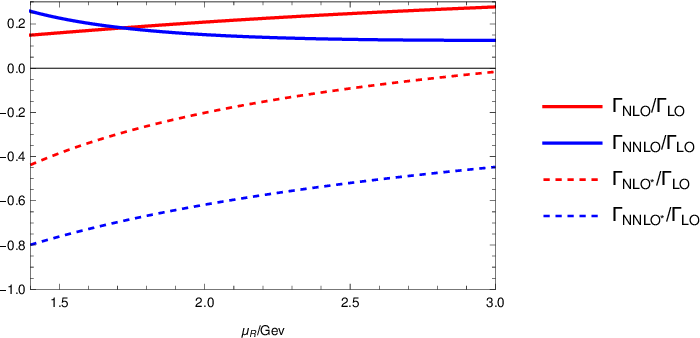}
		\caption[Renormalization scale dependence]{Renormalization scale dependence of partial decay widths {for $J/\psi$}: amplitude expansion (solid lines) vs. traditional expansion (dashed lines).}
		\label{fig:pic}
	\end{figure}

	Figure \ref{fig:pic} shows that amplitude expansion yields much better scale dependence than traditional expansion, although the NNLO result does not improve over NLO.
	
\begin{figure}[h]
		\centering
		\includegraphics[width=8cm, height=4cm]{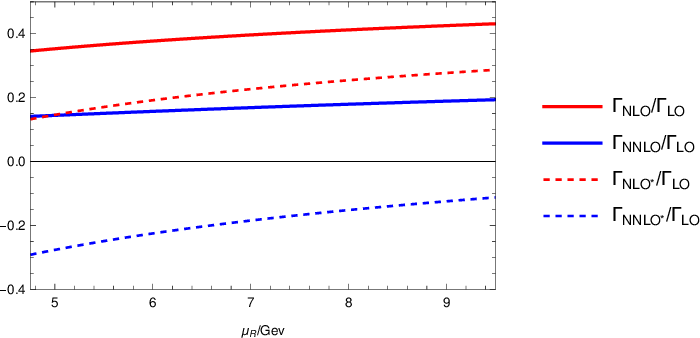}
		{\color{blue}\caption[Renormalization scale dependence]{Renormalization scale dependence of partial decay widths for $\Upsilon$: amplitude expansion (solid lines) vs. traditional expansion (dashed lines).}
		\label{fig:pic_upsilon}}
	\end{figure}

		  {\color{blue}Same analysis is extended to $\Upsilon\rightarrow 3 \gamma$ decay, as shown in Figure~\ref{fig:pic_upsilon}.  The amplitude expansion yields a much weaker scale dependence than the traditional expansion. For $\Upsilon$, which has a smaller $\alpha_s$ than $J/\psi$, the traditional NNLO result becomes negative over the entire scale range. Only the amplitude expansion produces a positive result.}

	With the given inputs for $J/\psi$, we obtain 
	\begin{align}
		\Gamma_{NLO}/\Gamma_{LO}=0.2771,
		\Gamma_{NNLO}/\Gamma_{LO}=0.1273. \nonumber 
	\end{align}

	Higher-order terms thus have decreasing effects, showing better convergence than the traditional expansion which gives negative results at both NLO and NNLO (Fig.\ref{fig:pic}).
	
	Theoretical uncertainties are assessed by independently varying key parameters over conventional ranges: renormalization scale $\mu_R\in[m_c, 2m_c]$, charm mass $m_c\in[1.4, 1.5]\,\text{GeV}$, NRQCD factorization scale $\mu_{\Lambda}\in[\frac{1}{2} m_c , m_c]$, and wave function at origin $|R(0)|^2\in[0.810, 1.454]\,\text{GeV}^3$. The results, with uncertainties decomposed by source, are in the first row of Table \ref{tab:error_budget}.
	
	For a complete prediction comparable with experiment, we include the known relativistic correction $\Gamma_{\text{Rel}}=\Gamma_{LO}[\frac{132-19\pi^2}{12(\pi^2-9)}+(\frac{16}{9}\ln \frac{\mu_{\Lambda}^2}{m_c^2}+68.913)\frac{\alpha_s}{\pi}]\langle v^2\rangle$ from Ref.~\cite{Feng:2012by}. Its central value and uncertainties from the same parameter variations are in the second row of Table \ref{tab:error_budget}. Our final prediction is the direct sum of these two contributions, listed in the third row.
	
	\begin{table}[tb]
		\renewcommand{\arraystretch}{1.5}
		\centering
		\renewcommand{\arraystretch}{1.5}
		\resizebox{1\columnwidth}{!}{
			\begin{tabular}{llcccc}
				\hline
				\multicolumn{6}{c}{\textbf{Theoretical predictions}} \\
				\hline
				Contribution & Central Value(eV) & $\Delta(\mu_R,m_c,\mu_\Lambda)$ & $\Delta|R(0)|^2$ & & \\
				\hline
				$\Gamma_{NNLO}(J/\psi)$ & 0.6659 & ${}^{+0.6921}_{-0}$ & ${}^{+0.3421}_{-0.08840}$ & &\\
				$\Gamma_{Rel}(J/\psi)$~\cite{Feng:2012by} & 0.2970 & ${}^{+3.304}_{-0.2346}$ & ${}^{+0.1653}_{-0.03943}$ & & \\
				\hline
				$\Gamma_{\text{Full}}(J/\psi)$ & 0.9629 & ${}^{+4.293}_{-0.004623}$ & ${}^{+0.5074}_{-0.1278}$ &&\\
				\hline
				BESIII (2013)~\cite{BESIII:2012lxx} & \multicolumn{5}{l}{$1.046 \pm 0.167 \pm 0.185 \quad (J/\psi)$} \\
				\hline
				$\Gamma_{NNLO}(\Upsilon)$ & 0.01295 & ${}^{+0.003114}_{-0.003252}$ &&&\\
				$\Gamma_{Rel}(\Upsilon)$~\cite{Sang:2020zdv} & -0.004342 & ${}^{+0.002652}_{-0.0004521}$  &&& \\
				\hline
				$\Gamma_{\text{Full}}(\Upsilon)$ & 0.008608 & ${}^{+0.002800}_{-0.0005995}$ &&& \\
				\hline
			\end{tabular}
		}
		\caption{Theoretical predictions and comparison with experiment for $\Gamma(J/\psi \to 3\gamma)$ and $\Gamma(\Upsilon \to 3\gamma)$ with uncertainties from independent variations of $\mu_R$, $m_c$, $\mu_{\Lambda}$, and $|R(0)|^2$. All values in eV.}
		\label{tab:error_budget}
	\end{table}
	
	The third and the fourth rows of Table \ref{tab:error_budget} compares our predictions with experiment. The pure NNLO result lies about $1.3\sigma$ below the central BESIII value. Adding the $\Gamma_{Rel}$ correction shifts the full prediction to $\Gamma_{\text{Full}} = 0.96 \, \text{eV}$, bringing it closer to the BESIII measurement and showing significantly better agreement. This indicates that both the complete NNLO QCD correction and the relativistic effect are essential for an accurate description. The dominant remaining uncertainty comes from $\mu_R$ variation, suggesting the potential impact of even higher perturbative orders.
	
	For $\Upsilon \to 3\gamma$, by choosing $m_c=0$, $n_l=4$,  $m_b=4.75\,\text{GeV}$, $m_{\Upsilon}=9.46\,\text{GeV}$, $|R(0)|^2=6.477\,\text{GeV}^3$ from \cite{Eichten:1995ch}, $\mu_R=2m_b$,  $\mu_{\Lambda}=m_b$ and $\alpha_s^{(3)}(\mu_R)=0.18055$, the results are obtained and presented at Table \ref{tab:error_budget}.
	
	We have presented the first complete NNLO QCD correction to $J/\psi \to 3\gamma$ within NRQCD factorization. The traditional expansion in Eq.(1) again becomes large and negative at NNLO, whereas without this new term the result agreed with experiment. To resolve the long-standing problem of negative decay and production rates~\cite{Caswell:1976nx, Mackenzie:1981sf, Keung:1982jb, Gong:2008ce}, we propose amplitude-level NRQCD factorization as in Eq.(\ref{eqn:factornization}) as a systematic prescription. The key distinction from the conventional approach is the reorganization of the perturbative series: by expanding at the amplitude level before squaring, we effectively regroup the terms in the series, which substantially improves convergence.

The resulting $\Gamma_{NNLO}(J/\psi \to 3\gamma)$ shows much better renormalization scale dependence and convergence than the traditional expansion (Fig.\ref{fig:pic}), consistent with the findings of our previous work~\cite{Huang:2023pmn}. It will be interesting to compare the convergence behavior of perturbative expansions between our prescription and the conventional approach in various exclusive processes where such a comparison is feasible.

{\color{blue}The improved convergence of our amplitude-level prescription can be understood from a fundamental quantum field theory perspective. The scattering amplitude is obtained from Feynman diagrams, and its modulus squared yields the physical decay or production probability. For inclusive processes, the cancellation of infrared (IR) divergences requires the squared amplitude to be expanded and truncated at a fixed order, which necessarily sacrifices the full modulus square. Historically, this same treatment was carried over to exclusive processes as well. For exclusive processes, however, all IR divergences cancel already at the amplitude level. One can therefore take the complete modulus squared of the full amplitude, which is precisely the proper definition of the probability distribution and guarantees its positivity. Hence, the amplitude-level prescription is the natural and physically motivated scheme from first principles for exclusive processes.}


{\color{blue}
Improving the convergence of QCD perturbative series has long been a central theme in the field. Our amplitude-level prescription is complementary to existing methods: while the BLM approach~\cite{BLM1983} and PMC~\cite{PMC2013} optimize the renormalization scale choice to absorb non-conformal $\beta$-function terms, and resummation techniques~\cite{Sang:2023liy} address specific classes of large contributions, our method operates at the level of organizing the perturbative expansion itself---expanding at the amplitude rather than the cross-section level. These approaches can be combined: one could apply BLM/PMC scale optimization on top of the amplitude-level expansion to further reduce scale uncertainty. The advantage of amplitude-level factorization is that it directly addresses the structural origin of the negative-rate problem at its root, without introducing additional scale-setting procedures. However, it is limited to exclusive processes where IR cancellations are fully contained at the amplitude level.
}	
	
	Our final prediction, $\Gamma(J/\psi \to 3\gamma) = 0.96^{+4.32}_{-0.13} \, \text{eV}$, combining the NNLO contribution with the known relativistic correction \cite{Feng:2012by} by direct addition, shows markedly improved agreement with the BESIII measurement ($1.046 \pm 0.167 \pm 0.185\, \text{eV}$). 
	In the same way, $\Gamma(\Upsilon \to 3\gamma) = 0.0086^{+0.0028}_{-0.0006} \, \text{eV}$ is obtained and the uncertainties are clearly much smaller. 
	{\color{blue}Meanwhile, the uncertainty may be further reduced if above-mentioned scale optimization is applied}
	
	The remaining $\sim 9\%$ central value discrepancy and the large theoretical uncertainty are mainly due to renormalization scale dependence and the additive treatment of $v^2$ and $\alpha_s v^2$ contributions. In our systematic prescription, these terms should be added at the amplitude factorization level, which is beyond the scope of this work and could be addressed in future studies.
	
	Nevertheless, our work establishes a crucial new benchmark. It demonstrates the feasibility of complete NNLO calculations for quarkonium electromagnetic decays, 40 years after the NLO calculation, and provides a solid foundation for further improvements. Reducing the dominant scale uncertainty will require N$^3$LO corrections or resummation techniques. Our results also offer valuable input for ongoing and future high-precision experiments such as BESIII and the planned Super Tau-Charm Factory, enabling more stringent tests of QCD in the heavy-quark sector.
	
	The author J. X. Wang thanks Prof. H. N. Li and Prof. C. D. Lu for discussions on amplitude-level factorization. This work is supported by the National Natural Science Foundation of China with Grants No. 12135013.
	
	\bibliography{references}

\begin{thebibliography}{0}%
\makeatletter
\providecommand \@ifxundefined [1]{%
 \@ifx{#1\undefined}
}%
\providecommand \@ifnum [1]{%
 \ifnum #1\expandafter \@firstoftwo
 \else \expandafter \@secondoftwo
 \fi
}%
\providecommand \@ifx [1]{%
 \ifx #1\expandafter \@firstoftwo
 \else \expandafter \@secondoftwo
 \fi
}%
\providecommand \natexlab [1]{#1}%
\providecommand \enquote  [1]{``#1''}%
\providecommand \bibnamefont  [1]{#1}%
\providecommand \bibfnamefont [1]{#1}%
\providecommand \citenamefont [1]{#1}%
\providecommand \href@noop [0]{\@secondoftwo}%
\providecommand \href [0]{\begingroup \@sanitize@url \@href}%
\providecommand \@href[1]{\@@startlink{#1}\@@href}%
\providecommand \@@href[1]{\endgroup#1\@@endlink}%
\providecommand \@sanitize@url [0]{\catcode `\\12\catcode `\$12\catcode
  `\&12\catcode `\#12\catcode `\^12\catcode `\_12\catcode `\%12\relax}%
\providecommand \@@startlink[1]{}%
\providecommand \@@endlink[0]{}%
\providecommand \url  [0]{\begingroup\@sanitize@url \@url }%
\providecommand \@url [1]{\endgroup\@href {#1}{\urlprefix }}%
\providecommand \urlprefix  [0]{URL }%
\providecommand \Eprint [0]{\href }%
\providecommand \doibase [0]{https://doi.org/}%
\providecommand \selectlanguage [0]{\@gobble}%
\providecommand \bibinfo  [0]{\@secondoftwo}%
\providecommand \bibfield  [0]{\@secondoftwo}%
\providecommand \translation [1]{[#1]}%
\providecommand \BibitemOpen [0]{}%
\providecommand \bibitemStop [0]{}%
\providecommand \bibitemNoStop [0]{.\EOS\space}%
\providecommand \EOS [0]{\spacefactor3000\relax}%
\providecommand \BibitemShut  [1]{\csname bibitem#1\endcsname}%
\let\auto@bib@innerbib\@empty
\end{thebibliography}%


\begin{thebibliography}{99}
		\bibitem{Appelquist:1975} T.~Appelquist and H.~D.~Politzer, Phys. Rev. Lett. \textbf{34}, 43 (1975).
		\bibitem{DeRujula:1975} A.~De Rujula and S.~L.~Glashow, Phys. Rev. Lett. \textbf{34}, 46 (1975).
		\bibitem{Bodwin:1994jh}
		G.~T.~Bodwin, E.~Braaten, and G.~P.~Lepage,
		Phys. Rev. D \textbf{51} (1995), 1125-1171
		[erratum: Phys. Rev. D \textbf{55} (1997), 5853].
		\bibitem{Caswell:1986} W.~E.~Caswell and G.~P.~Lepage, Phys. Lett. B \textbf{167}, 437 (1986).
		\bibitem{Caswell:1976nx}
		W.~E.~Caswell, G.~P.~Lepage, and J.~R.~Sapirstein,Phys. Rev. Lett. \textbf{38} (1977), 488.
		\bibitem{Adkins:1996} G.~S.~Adkins, Phys. Rev. Lett. \textbf{76}, 4903 (1996).
		\bibitem{CLEO:2008qfy}
		G.~S.~Adams \textit{et al.} [CLEO],Phys. Rev. Lett. \textbf{101} (2008), 101801.
		\bibitem{BESIII:2012lxx}
		M.~Ablikim \textit{et al.} [BESIII],Phys. Rev. D \textbf{87} (2013) no.3, 032003.
		\bibitem{Mackenzie:1981sf}
		P.~B.~Mackenzie and G.~P.~Lepage,Phys. Rev. Lett. \textbf{47} (1981), 1244.
		\bibitem{Keung:1982jb}
		W.~Y.~Keung and I.~J.~Muzinich,Phys. Rev. D \textbf{27} (1983), 1518.
		\bibitem{Braaten:1998au}
		E.~Braaten and Y.~Q.~Chen,Phys. Rev. D \textbf{57} (1998), 4236-4253
		[erratum: Phys. Rev. D \textbf{59} (1999), 079901].
		\bibitem{Bodwin:1998mn}
		G.~T.~Bodwin and Y.~Q.~Chen,Phys. Rev. D \textbf{60} (1999), 054008.
		\bibitem{Feng:2012by}
		F.~Feng, Y.~Jia, and W.~L.~Sang,Phys. Rev. D \textbf{87} (2013) no.5, 051501.	
		\bibitem{Sang:2020zdv}
		W.~L.~Sang, F.~Feng, and Y.~Jia,Phys. Rev. D \textbf{102} (2020) no.9, 094021.
		\bibitem{Meng:2019lkt}
		Y.~Meng, C.~Liu, and K.~L.~Zhang,Phys. Rev. D \textbf{102}, no.5, 054506 (2020).	
		\bibitem{Gong:2008ce}
		B.~Gong and J.~X.~Wang,Phys. Rev. Lett. \textbf{100} (2008), 181803.
		\bibitem{Sang:2023liy}
		W.~L.~Sang, F.~Feng, Y.~Jia, Z.~Mo, J.~Pan, and J.~Y.~Zhang,Phys. Rev. Lett. \textbf{131} (2023) no.16, 161904.
		\bibitem{Huang:2023pmn}
		X.~D.~Huang, B.~Gong, R.~C.~Niu, H.~M.~Yu, and J.~X.~Wang,JHEP \textbf{02} (2024), 055.
		\bibitem{Beneke:1997jm}
		M.~Beneke, A.~Signer, and V.~A.~Smirnov,Phys. Rev. Lett. \textbf{80} (1998), 2535-2538.
		\bibitem{Kniehl:2006qw}
		B.~A.~Kniehl, A.~Onishchenko, J.~H.~Piclum, and M.~Steinhauser,Phys. Lett. B \textbf{638} (2006), 209-213.
		\bibitem{Adkins:1996odo}
		G.~S.~Adkins,Phys. Rev. Lett. \textbf{76} (1996), 4903-4906.
		\bibitem{Hahn:2000kx}
		T.~Hahn, Comput.\ Phys.\ Commun.\ \textbf{140}, 418 (2001).
		\bibitem{Calcloop}
		\texttt{CalcLoop} is a Mathematica package developed by Yan-Qing Ma, available at https://gitlab.com/multiloop-pku/calcloop.
		\bibitem{Kira:2021}
		J.~Klappert, F.~Lange, P.~Maierh\"ofer, and J.~Usovitsch,Comput.\ Phys.\ Commun.\ \textbf{266}, 108024 (2021).
		\bibitem{Liu:2022chg}
		X.~Liu and Y.-Q.~Ma, Comput.\ Phys.\ Commun.\ \textbf{283}, 108565 (2023).
		\bibitem{Campbell:2007ws}
	    Campbell, John M. and Maltoni, F. and Tramontano, F.,
		Phys. Rev. Lett. \textbf{98}, (2007), 252002.
		\bibitem{Liu:2020kpc}
		X.~Liu, Y.~Q.~Ma, W.~Tao and P.~Zhang,
		Chin. Phys. C \textbf{45} (2021) no.1, 013115.
		\bibitem{Bodwin:2008}
		G.~T.~Bodwin, H.~S.~Chung, D.~Kang, J.~Lee, and C.~Yu,Phys.\ Rev.\ D \textbf{77}, 094017 (2008).
		\bibitem{RunDec:2017}
		F.~Herren and M.~Steinhauser,Comput.\ Phys.\ Commun.\ \textbf{224}, 333 (2018).
		\bibitem{Eichten:1995ch}
		E.~J.~Eichten and C.~Quigg,
		Phys. Rev. D \textbf{52} (1995), 1726-1728.
		\bibitem{BLM1983}
		S.J. Brodsky, G.P. Lepage and P.B. Mackenzie, Phys.
		Rev. D 28 228 1983, G. Grunberg and A. L. Kataev,
		Phys. Lett. B 279 352-358 (1992), S.J. Brodsky and
		H.J. Lu, Phys. Rev. D 51 3652 1995.
		\bibitem{PMC2013}
		M.~Mojaza, S.~J.~Brodsky and X.~G.~Wu,
		Phys. Rev. Lett. \textbf{110} (2013), 192001;
		S.~J.~Brodsky, M.~Mojaza and X.~G.~Wu,
		Phys. Rev. D \textbf{89} (2014), 014027;
		X.~G.~Wu, S.~J.~Brodsky and M.~Mojaza,
		Prog. Part. Nucl. Phys. \textbf{72} (2013), 44-98.
	\end{thebibliography}
	
\end{document}